# Analytical Description of Mixed Ohmic and Space-Charge-Limited Conduction in Single-Carrier Devices

Jason A. Röhr[1,*], Roderick C. I. MacKenzie[2]

[1]*Department of Chemical and Biomolecular Engineering, New York University, Brooklyn, 11201 NY, United States of America*
[2]*Faculty of Engineering, University of Nottingham, Nottingham, NG7 2RD, United Kingdom*

While space-charge-limited current measurements are often used to characterize charge-transport in relatively intrinsic, low-mobility semiconductors, it is currently difficult to characterize lightly or heavily doped semiconductors with this method. By combining the theories describing ohmic and space-charge-limited conduction, we derive a general analytical approach to extract the charge-carrier density, the conduction-band edge and the drift components of the current density-voltage curves of a single-carrier device when the semiconductor is either undoped, lightly doped or heavily doped. The presented model covers the entire voltage range, i.e., both the low-voltage regime and the Mott-Gurney regime. We demonstrate that there is an upper limit to how doped a device must be before the current density-voltage curves are significantly affected, and we show that the background charge-carrier density must be considered to accurately model the drift component in the low-voltage regime, regardless of whether the device is doped or not. We expect that the final analytical expressions presented herein to be directly useful to experimentalists studying charge transport in novel materials and devices.

*jasonrohr@nyu.edu

## I. INTRODUCTION

Space-charge-limited current (SCLC) measurements rely on the interpretation of data obtained from single-carrier devices where only one charge-carrier type (e.g. electrons) dominates the current flow (**Fig. 1a**), and are amongst the most commonly used methods for determining charge-carrier mobilities, $\mu$, of relatively intrinsic semiconductors.[1–7] SCLC measurements are highly popular due to the fact that: i) The single-carrier devices used for SCLC measurements are relatively easy to fabricate and operate under similar conditions to that of optoelectronic devices; ii) fabricating single-carrier devices does not require a large amount of material, which is beneficial when newly-developed semiconductors are being probed where material is scarce; iii) SCLC measurements are relatively easy to perform and do not require access to powerful magnets or lasers; iv) charge transport of electrons and holes can be probed separately by an appropriate choice of contacts, and; v) SCLC measurements can yield information about energetic disorder, doping and traps if proper models are used to interpret the results. SCLC measurements have therefore become a standard method to characterize a wide variety of novel semiconductors, such as metal chalcogenides,[8] amorphous silicon,[9] organic semiconductors,[10–12] fullerenes,[13,14] and metal-halide perovskites.[15,16]

To obtain charge-transport information from SCLC measurements, one must fit a model to the experimental current density-voltage (*J-V*) curves. Several analytical models have previously been proposed that describe intrinsic semiconductors with relatively high accuracy;[1,17–19] however, semiconductors typically contain defects that can give rise to doping and traps. These defects can significantly affect the measured *J-V* curves, and it is therefore important to utilize a model that can account for said defects to obtain reliable charge-transport characteristics.[20–22] Although a number of numerical models have been developed that can account for defects of various kind, [22–29] analytical models are easier to employ and are therefore more often used by experimentalists. It is therefore important to develop accurate analytical models that can aid in describing the situations where the semiconductor is not intrinsic.[24,25] A number of analytical models have been developed to account for non-ideal semiconductors, such as when the semiconductor contains traps described by exponential tails,[20,24] when Poole-Frenkel effects dominate,[21,30] or when charge transport is limited by Gaussian disorder;[31] however, an accurate analytical model that describes the situation where the semiconductor is doped, does not exist.

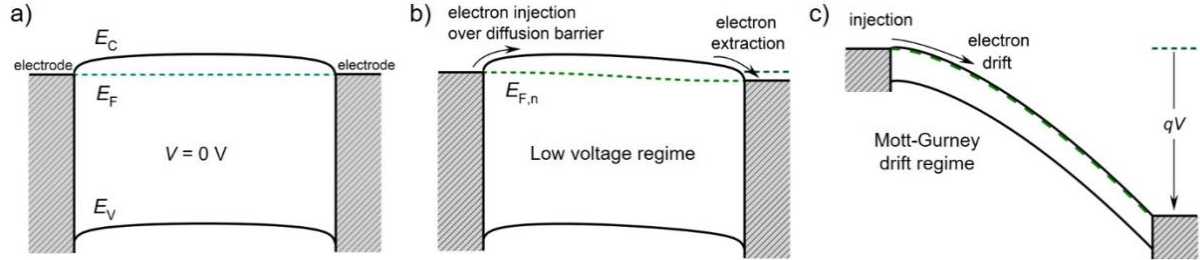

**Figure 1** – Schematic of the energy levels of a symmetric electron-only single-carrier device when operated under different applied voltages: **a)** 0 V where no current is flowing, **b)** low voltage, in which a linear *J-V* relationship is commonly observed (typically for *V* < 0.9 V), **c)** when enough voltage is applied so that the current has transitioned into the Mott-Gurney regime and a relationship close to $J \propto V^2$ is observed (for *V* > 0.9 V). $E_C$ and $E_V$ are the conduction- and valence-band edges, $E_F$ is the Fermi level at *V* = 0 V and $E_{F,n}$ is the electron quasi-Fermi level at *V* > 0 V.

As a voltage is applied across a single-carrier device, the charge-transport characteristics typically transition between regimes at low and high voltage. For a trap- and doping-free semiconductor, the current in the low-voltage regime is typically not dominated by thermally-generated intrinsic charge carriers, $n_i$, but rather due to the background charge carriers, $n_b$, injected into the single-carrier device from the contacts during Fermi-level equilibration.[32] This means that $n_b$ far exceeds $n_i$, and it has been shown that the current obtained from an electron-only device due to these charge carriers can be accurately described by,[17,18]

$$J = 4\pi^2 \frac{k_B T}{q} \mu_n \varepsilon_r \varepsilon_0 \frac{V}{L^3} \qquad (1)$$

where $k_B T$ is the thermal energy, $q$ is the elementary charge, $\mu_n$ is the electron mobility, $\varepsilon_r \varepsilon_0$ is the permittivity, $V$ is the applied voltage and $L$ is the thickness of the semiconductor. The energy levels for an electron-only device operated under the low-voltage conditions resulting in the *J-V* behavior described by eq. 1 are shown in **Fig. 1b**. In the case where a hole-only device is being measured, $\mu_n$ is replaced by the hole mobility, $\mu_p$.

When enough voltage is applied to ensure that the current flow is fully dominated by drift (**Fig. 1c**), the *J-V* curves can be modelled by the classical Mott-Gurney square law,[1]

$$J = \frac{9}{8} \mu_n \varepsilon_r \varepsilon_0 \frac{V^2}{L^3}. \qquad (2)$$

Despite its inability to describe doped semiconductors or semiconductors with trap states, the Mott-Gurney law is the most commonly used analytical model for characterizing SCLC data. Given that the semiconductor is free from traps and doping, and the contacts for injection and extraction are perfectly ohmic, and given that $\varepsilon_r$ and $L$ are known (and that $L$ is not too small), eq. 1 and 2 can be fitted to an SCLC *J-V* curve to extract $\mu_n$ as the only unknown quantity.[32,33] These two equations (eq. 1 and 2) combined therefore give an excellent description of the *J-V* curves obtained from SCLC measurements when the measured semiconductor is perfectly intrinsic and the single-carrier device does not suffer from injection issues at the contacts (see **Fig. 2a**).

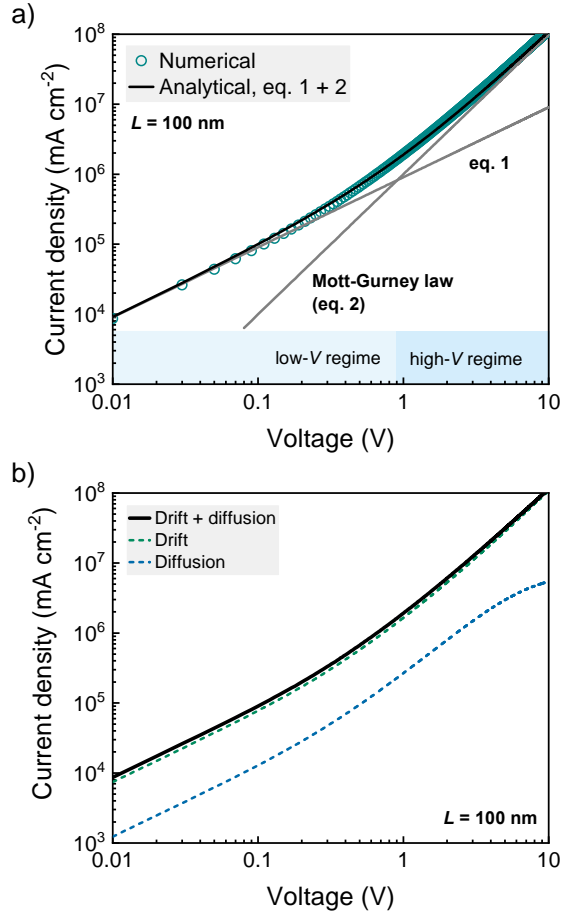

**Figure 2** – **a)** Comparison of a numerically calculated, undoped single-carrier device with ohmic contacts (circles) with fits with eq. 1 and the Mott-Gurney law, eq. 2 (solid grey lines) along with the sum of eq. 1 and eq. 2 (solid black line). The low- and high-voltage regimes are denoted. **b)** Total current density (drift + diffusion, solid line) and the individual contributions to the total current due to drift and diffusion (dashed lines).

While one must assume that the semiconductor does not contain defects in deriving eq. 1 and 2, it has been identified that SCLC measurements themselves could potentially be used to characterize lightly doped semiconductors if proper models are employed.[34] To this end, SCLC measurements have so far been used to characterize lightly doped organic semiconductors by assuming that the mobility can be obtained from the Mott-Gurney law in the high-voltage regime, and then employing Ohm's law to estimate the conductivity in the low-voltage regime,[35]

$$J = q\mu_\mathrm{n} n_\mathrm{D} \frac{V}{L} \quad (3)$$

where $n_\mathrm{D}$ is the free electron density due to added donors. With the knowledge of $\mu_\mathrm{n}$ from the Mott-Gurney law, $n_\mathrm{D}$ can now, in principle, be determined as the only unknown quantity in eq. 3. It should therefore be possible to yield information about both $\mu_\mathrm{n}$ and $n_\mathrm{D}$ of a lightly doped device; however, as the semiconductor becomes increasingly doped, the current across the entire voltage range becomes increasingly ohmic, and therefore less space-charge limited, and it will no longer be possible to fit the Mott-Gurney law in the drift regime to extract $\mu_n$. One is therefore left with the $\mu_n n_D$ product, and neither $\mu_n$ nor $n_D$, separately.[32] Since SCLC measurements are so commonly used, and since scientist are employing such measurements to quantify doping, it is important to develop simple methods that can reliably extract $\mu_n$ and $n_D$ from the *J-V* curves of both lightly doped and heavily doped semiconductors in single-carrier devices.

Using numerical calculations, we show that a generalized expression for SCLC in a undoped single-carrier device can be written as the sum of eq. 1 and the Mott-Gurney law (eq. 2). We also show that while the sum of eq. 2 and Ohm's law (eq. 3) is sufficient for describing the *J-V* curves of a single-carrier device where the semiconductor is heavily doped (in which case eq. 3 will dominate), it is not sufficient for describing a lightly doped semiconductor. We show that a series of analytical expressions can be derived that can describe the charge-carrier density, conduction-band edge and drift-current density of single-carrier devices regardless of whether the semiconductor is undoped, lightly doped or heavily doped. We present a simple condition for how doped the semiconductor must be before the *J-V* curves are significantly affected, and we show that to accurately model the *J-V* curves obtained from a lightly doped semiconductor, $n_b$ and $n_D$ must both be taken into account whereas $n_b$ can be ignored in the high-doping limit. The analytical expressions presented herein can be fitted to SCLC data to yield $\mu_n$ and $n_D$, simultaneously.

## II. NUMERICAL METHODS

To test the validity of our derived analytical expressions, developed in the next section, we compare them to numerical calculations of single-carrier devices.[36–38] This allows us to understand the validity of these expressions over a wide range of doping densities, while ensuring that certain semiconductor characteristics, that are commonly present in real semiconductors and single-carrier devices, such as traps and injection-barrier heights, could be omitted, while certain characteristics could be held constant, such as the mobility and thickness. This approach allowed for an elegant comparison between the derived analytical expressions with a type of numerical model that has been used to successfully analyze experimental data from both single-carrier devices and solar cells on several occasions.[26,39,40]

Our numerical model, *general-purpose photovoltaic device model* (gpvdm)[26,41] solves the drift-diffusion equations for electrons and holes,

$$J_n(x) = qn(x)\mu_n F(x) + qD_n \frac{dn(x)}{dx} \quad (4)$$

$$J_p(x) = qp(x)\mu_p F(x) - qD_p \frac{dp(x)}{dx} \quad (5)$$

to describe the movement of charge carriers, and Poisson's equation to describe the electrostatic potential,

$$\varepsilon_0 \nabla \varepsilon_r \nabla \varphi(x) = -\rho(x). \quad (6)$$

where $D_{n/p}$ are the Einstein-Smoluchowski diffusion coefficients, $n$ and $p$ are the total free electron and hole densities, $F$ is the electric field, $\varphi$ is the electric potential, and $\rho$ is the total charge density (accounting for all charge, both free and stationary).

The boundary conditions for the simulations were set by the interface charge-carrier density, $n_{int}$, via the injection-barrier heights, $q\phi_{inj}$, at the semiconductor-conductor interfaces at $x = 0$ and $x = L$,

$$n_{int}(x) = N_C \exp\left(-\frac{q\phi_{inj}(x)}{k_B T}\right). \quad (7)$$

Ohmic contacts were assumed for all the analytical derivations, $q\phi_{\text{inj}}(0) = 0$ eV and $q\phi_{\text{inj}}(L) = 0$ eV; however, the case where the contacts are non-ohmic was also considered to assess when considerable deviation could be expected from fitting with the final expressions.

The single-carrier devices were calculated using device parameters and materials constants chosen to represent a trap-free semiconductor/insulator: $E_g = 3$ eV, $N_C = N_V = 10^{20}$ cm$^{-3}$, $\mu_n = \mu_p = 1$ cm$^2$ V$^{-1}$s$^{-1}$, $\varepsilon_r = 10$ and $T = 300$ K (room temperature). Charge-carrier mobilities vary greatly between semiconductors, with measured mobilities in the range of $10^{-6}$ to $10^3$ cm$^2$ V$^{-1}$s$^{-1}$. The mobility chosen for the simulations, $1$ cm$^2$ V$^{-1}$s$^{-1}$, is close to what is commonly observed for amorphous and microcrystalline silicon[42] along with various solution-processed metal chalcogenides.[43] The magnitude of the charge-carrier mobility only acts to increase the overall magnitude of the current density, as seen in eq. 4 and 5, and does not affect shape of the *J-V* curves (**Fig. S1**). Therefore, any value of the charge-carrier mobility could have been chosen, and the presented results would not be affected.

For the sake of simplicity, we will only consider electron-only devices. All calculations are analogous for the cases where hole-only devices are considered. The semiconductor was doped with electrons by adding a uniform distribution of positive space charge to the semiconductor, $N_D$, giving rise to an additional electron doping density, $n_D$.

## III. RESULTS & DISCUSSION

In this section we arrive at an expression describing *J-V* curves, across both the low- and high-voltage regimes, of undoped, lightly doped or heavily doped single-carrier devices. To arrive at this expression, we initially derive an expression that describes the *J-V* characteristics of an undoped device and then extend that expression to cover doped devices. In part A we determine that both the low- and high-voltage regimes are dominated by drift. In part B we derive an expression for the charge-carrier density of an undoped device as a function of position and voltage, and we then use this to derive an expression for the *J-V* characteristics across both voltage regime (part C). In part D we derive an expression for the charge-carrier density as a function of position and voltage for a doped device, and we then arrive at *J-V* expression for a doped device (part E). Finally, in part F, we derive an expression for the condition for when doping dominates the *J-V* curves rather than the background charge-carrier density due to injection from the electrodes during Fermi-level equilibration.

### A. Drift and diffusion currents from an undoped device

An electron-only single-carrier device is achieved by matching both contact work functions with the conduction-band edge of the semiconductor, $E_C$, to achieve ohmic contacts, as shown in **Fig. 1a**. In the case that both contact work functions align with $E_C$, the device is called symmetric and the *J-V* curves are expected to be similar regardless of whether a positive or negative bias is applied.[37] A numerically calculated, representative *J-V* curve of an undoped, 100 nm semiconductor single-carrier device, is shown in **Fig. 2a**. A transition from a linear *J-V* behavior at low voltage to a $J \propto V^2$ behaviour at higher voltage is observed, which is expected from a symmetric single-carrier device. These transport regimes correspond to the scenarios shown in **Fig. 1b,c** and can be described by eq. 1 and eq. 2 (the Mott-Gurney law), respectively (see **Fig. 2a**).

The contributions from drift and diffusion to the full *J-V* curve shown in **Fig. 2a** is shown in **Fig. 2b**, both as dashed lines, with the sum shown with the solid line (the implementation of the calculations of the drift and diffusion components are explained in the supplementary material). It is well-known that the Mott-Gurney regime is drift dominated; however, it is here shown that the low-voltage regime is also primarily dominated by drift, with the diffusion current being approximately one order of magnitude less than the contribution from the drift current. This can seem like a surprising result due to the inclusion of the diffusion coefficient in eq. 1 ($D = \mu k_B T/q$), but can be understood by the current being a drift current due to the background charge carriers being injected via diffusion during Fermi-level equilibration. Since the current is dominated by drift across all voltages, we can obtain a description for the current density as long as a description for the mean charge-carrier density, $\langle n \rangle$, valid for all voltages, can be obtained.

## B. Charge-carrier density and conduction-band edge for an undoped device

The total equilibrium electron density in an undoped electron-only device at $V = 0$ V can be described by the sum of the intrinsic electron density, $n_i$, and the background electron density due to injection from the contacts during Fermi-level equilibration, $n_b$: $n = n_i + n_b$. As discussed below, while $n_i$ can be ignored for cases when a semiconductor with a relatively large band gap is being measured, $n_b$ is very large for relatively thin devices. This is especially important at the interfaces between the semiconductor and the contacts, regardless of the magnitude of the band gap.[32] For the numerical calculation of a relatively thin (100 nm) single-carrier device, $n_b(x)$ is very large with the majority of the charge carriers residing near the semiconductor-contact interfaces (see **Fig. 3a**)[32]. For devices with larger thicknesses, $L = 1$ μm and $L = 10$ μm, the overall magnitude of $n_b(x)$ decreases with most of the charge carriers still residing near the interfaces (see **Fig. 3b,c**).

Simmons has shown that $n_b(x)$ can be written analytically as a function of position within the device as,[44,45]

$$n_b(x) = \frac{2\pi^2 \varepsilon_r \varepsilon_0 k_B T}{q^2 L^2} \left[\cos^2\left\{\frac{\pi x}{L} - \frac{\pi}{2}\right\}\right]^{-1}. \quad (8)$$

Charge-carrier density profiles with similar form to what is described by eq. 8 have been observed for amorphous silicon,[46] CIGS,[47,48] CdTe[49] and metal-halide perovskites.[50] In fact, an expression similar to the expression proposed by Simmons was recently used to characterize these profiles.[51]

As shown in **Fig. 3a-c**, the shape and magnitude of $n_b(x)$ can be accurately described with eq. 8 regardless of the thickness of the semiconducting layer. The observed overall decrease in $n_b(x)$ with increased $L$ can also be understood from eq. 8 as $n_b \propto L^{-2}$. For semiconductors with relatively large band gaps, $E_g > 2$ eV, the thickness of the semiconductor would have to be much larger than 10 μm before the intrinsic charge-carrier density will dominate. $n_i$ can therefore be ignored for most practical purposes and eq. 8 is therefore adequate for describing the charge-carrier density of an undoped semiconductor at 0 V. In cases where $n_i$ cannot be ignored, the electron density at $V = 0$ V will simply be equal to $n_b + n_i$.

As a voltage is applied across the electron-only device ($V > 0$ V), electrons are injected from the injecting contact into the semiconductor, increasing the electron density across the semiconductor, $n > n_b$. In the low-voltage regime, the charge-carrier density does not deviate from the equilibrium charge-carrier density at 0 V by an appreciable amount even when a small voltage is applied.[24] In fact, as can be seen in **Fig. 3a-c**, a significant voltage must be applied before a significant increase in the charge-carrier density is observed, with the distribution becoming asymmetric with a larger density near the vicinity of the injection metal-semiconductor interface at $x = 0$. To derive an analytical expression for this increase in the charge-carrier density as a voltage is applied, we combine the charge-carrier density in the Mott-Gurney regime, $n_{MG}$ with the background charge-carrier density to describe the charge-carrier density for $V > 0$ V, $n = n_b + n_{MG}$. The voltage-dependent charge-carrier density in the Mott-Gurney regime is given by,

$$n_{MG}(x,V) = \frac{3}{4}\frac{\varepsilon_r \varepsilon_0}{q}\frac{V}{L^{3/2}}x^{-1/2}. \quad (9)$$

and we can write the total charge-carrier density as a function of voltage as,

$$n(x,V) = \frac{2\pi^2 \varepsilon_r \varepsilon_0 k_B T}{q^2 L^2}\left[\cos^2\left\{\frac{\pi x}{L}-\frac{\pi}{2}\right\}\right]^{-1} + \frac{3}{4}\frac{\varepsilon_r \varepsilon_0}{q}\frac{V}{L^{3/2}}x^{-1/2}. \quad (10)$$

As seen in **Fig. 3a-c**, equation 10 is a good description of charge-carrier density at various thicknesses (100 nm, 1 um and 10 μm) and over a range of applied voltages (0, 1, 10 and 100 V), especially near the injection contact and in the middle of the device.

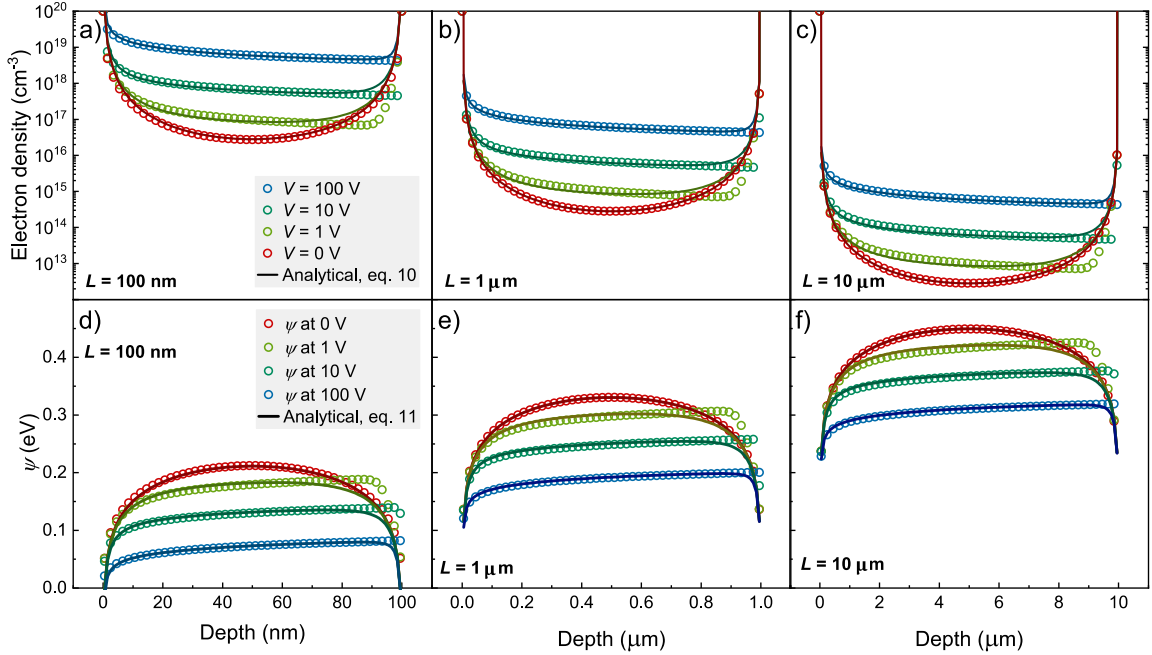

**Figure 3 – a-c)** Numerical calculations (circles) for the electron density at 0, 1, 10 and 100 V, for a 100 nm, 1 μm and 10 μm, respectively. Analytical calculations with eq. 10 are shown as solid lines. d-f) Equivalent calculations for $\psi = E_C - E_{F,n}$ and calculations with eq. 11.

Additionally, an analytical description can be written for the difference between the conduction-band edge, $E_C$ and the quasi-Fermi level, $E_{F,n}$, $\psi = E_C - E_{F,n}$, via $n = N_C \exp(-\{E_C - E_{F,n}\}/k_B T)$ as,

$$\psi(x,V) = -k_B T \ln\left(\frac{2\pi^2 \varepsilon_r \varepsilon_0 k_B T}{q^2 L^2 N_C}\left[\cos^2\left\{\frac{\pi x}{L}-\frac{\pi}{2}\right\}\right]^{-1} + \frac{3}{4}\frac{\varepsilon_r \varepsilon_0}{q N_C}\frac{V}{L^{3/2}}x^{-1/2}\right). \quad (11)$$

Just as eq. 10 is a good description for the charge-carrier density as a function of voltage (see **Fig. 3a-c**), equation 11 is a good description for the conduction-band edge, as seen in **Fig. 3d-f**.

## C. Total *J-V* description of an undoped single-carrier device

To arrive at an expression for the drift current density, $J = q\mu_n \langle n \rangle V/L$, the mean of the charge-carrier density, $\langle n \rangle$, must be calculated. The arithmetic mean of $n_b(x)$, $\langle n_b \rangle = L^{-1} \int_0^L n_b(x)\, dx$, cannot be calculated as the integral does not converge; however, $n_b(x)^{-1}$ can be integrated, and the harmonic mean can thus be calculated. To obtain an expression for the total drift current we therefore take the sum of the harmonic means, $\langle n \rangle = \langle n_b \rangle + \langle n_{MG} \rangle$,

$$\langle n \rangle = \frac{1}{\frac{1}{L}\int_0^L n_b^{-1}\, dx} + \frac{1}{\frac{1}{L}\int_0^L n_{MG}^{-1}\, dx} \qquad (12)$$

yielding,

$$\langle n \rangle = \frac{4\pi^2 \varepsilon_r \varepsilon_0 k_B T}{q^2 L^2} + \frac{9\varepsilon_r \varepsilon_0 V}{8qL^2}. \qquad (13)$$

Inserting eq. 13 into $J = q\mu_n \langle n \rangle V/L$, we obtain the total drift current density as the sum of eq. 1 and eq. 2 (the Mott-Gurney law),

$$J = 4\pi^2 \frac{k_B T}{q} \mu_n \varepsilon_r \varepsilon_0 \frac{V}{L^3} + \frac{9}{8} \mu_n \varepsilon_r \varepsilon_0 \frac{V^2}{L^3}. \qquad (14)$$

Since eq. 10 is a good description for the charge-carrier density as a function of voltage (see **Fig. 3a-c**), eq. 14 is likewise a good description for the total current density in the same voltage range (see **Fig. 2a**). Equation 14 can therefore be used to fit the entire *J-V* curve to extract the charge-carrier mobility when the semiconductor is undoped and ohmic contacts are achieved between the semiconductor and the contacts. That a generalized expression that covers both the low and high voltage regimes can be written as the sum of eq. 1 and 2 has, to our knowledge, not been presented in the literature.

## D. Charge-carrier density and conduction-band edge for a doped device

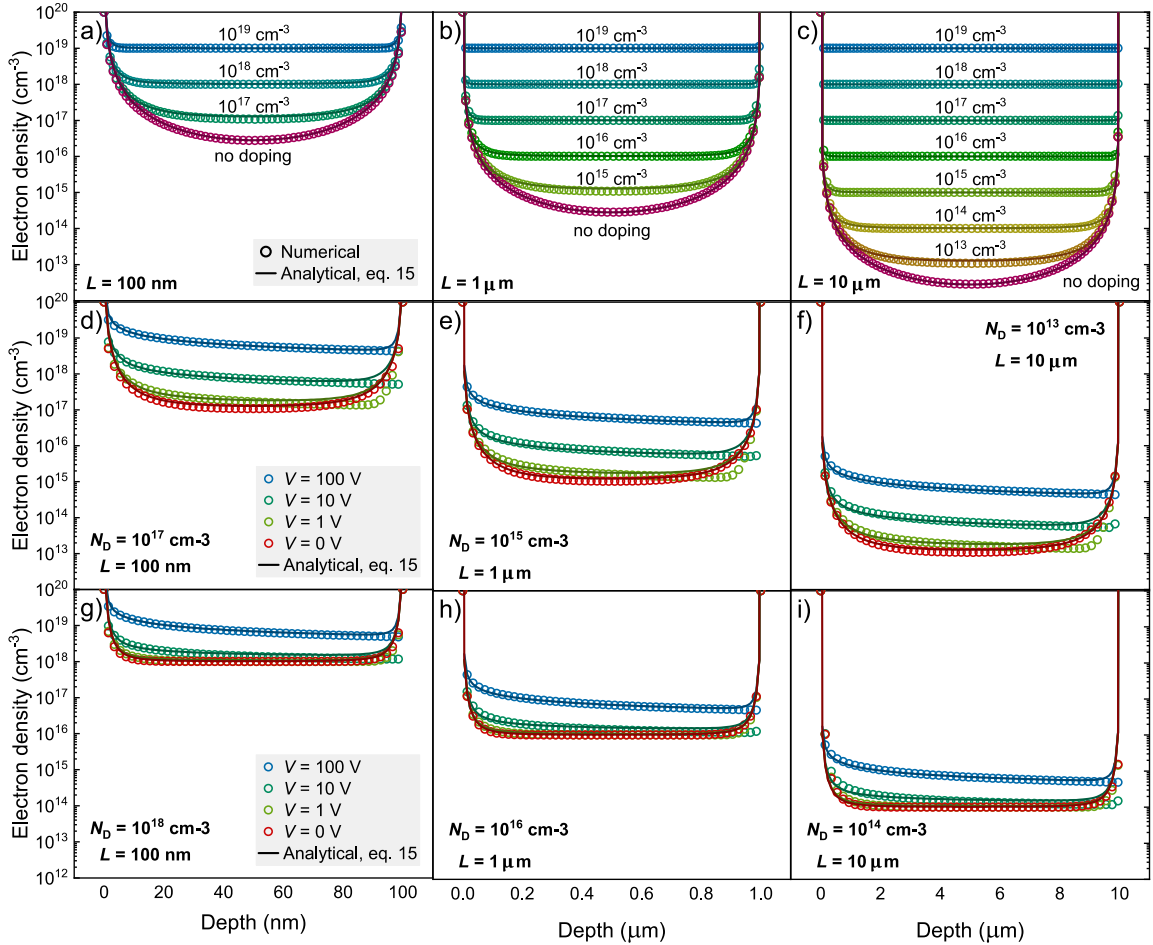

**Figure 4** – Analytically and numerically calculated electron densities of undoped and doped symmetric electron-only devices under various applied voltages. **a-c)** Electron densities for a 100 nm, 1 μm and 10 μm device when $N_D$ is varied from $10^{13}$-$10^{19}$ cm$^{-3}$. Hollow circles are the numerical values and solid lines are calculated using eq. 15 for $V = 0$ V. **d-i)** Calculated electron densities for doped 100 nm, 1 μm and 10 μm single-carrier devices at applied voltages in the range of 0 to 100 V, with circles being the numerical calculations and solid lines the analytical calculations with eq. 15.

When doping in the form of positively charged donors, $N_D$, is uniformly distributed throughout the semiconductor (as described in section II), the free electron density increases. Ignoring the intrinsic charge carriers, and assuming that each added donor is thermalized and gives rise to a free electron, $N_D = n_D$, the density of free electrons at $V = 0$ V is given by $n(x) = n_b(x) + n_D(x)$. Since $n_b(x)$ is very large at the semiconductor-metal interfaces, the background charge-carrier density will be a main contributor to the charge-carrier density even when $n_D$ is large and the semiconductor is doped close to degeneracy. Both $n_b$ and $n_D$ must therefore be taken into account when modelling the current density.

Numerically calculated electron densities from electron-only devices of various thicknesses, $L = 100$ nm-10 μm, both undoped and doped, $N_D = 10^{13}$–$10^{19}$ cm$^{-3}$, are shown in **Fig. 4a-c**. As doping is added to the 100 nm thick semiconductor, the electron density increases in the bulk of the device with the electron density at the boundaries still dominated by the background charge-carrier density (see **Fig. 4a**). Since $n_b$ is very large across the entire depth of the semiconductor when the device is relatively thin, a significant doping density must be incorporated before $n$ increases above the background density as the electron density will be entirely masked by $n_\text{b}$. For the modelled 100 nm device, a doping density of $>10^{16}$ cm$^{-3}$ must be added before $n$ increases by a significant amount above $n_\text{b}$. For the thicker devices, a lower doping density can be detected due to the decrease in $n_\text{b}$ (see **Fig. 4b,c**).

We can now write a full description of the electron density for a doped device as a sum $n_\text{b}(x)$, $n_\text{MG}(x,V)$ and $n_\text{D}(x)$,

$$n(x,V) = \frac{2\pi^2 \varepsilon_\text{r} \varepsilon_0 k_\text{B} T}{q^2 L^2}\left[\cos^2\left\{\frac{\pi x}{L} - \frac{\pi}{2}\right\}\right]^{-1} + \frac{3}{4}\frac{\varepsilon_\text{r}\varepsilon_0}{q}\frac{V}{L^{3/2}}x^{-1/2} + n_\text{D}(x). \tag{15}$$

**Figure 4a-c** shows that an excellent agreement between the numerical calculations and eq. 15 evaluated at $V = 0$ V is found, regardless of the thickness and donor density. It can be seen that it is particularly important to account for both $n_\text{b}$ and $n_\text{D}$ when the device is either thin or when the doping density is relatively low, as there is a significant amount of charge carriers at the interfaces that must be accounted for. For a thin device this is even true as the doping density tends towards degeneracy ($N_\text{D} \to N_\text{C}$). It should also be noted that while certain curves are labelled as undoped in **Fig. 4a-c**, eq. 15 will still give a good description for the electron density as the first term will outweigh the third term for low values of $n_\text{D}(x)$. As shown in **Fig. 4d-i**, as a voltage is applied across a doped device, eq. 15 is also a good description, regardless of the magnitude of the thickness or the doping density.

Similarly to what was calculated for the undoped devices, $\psi(x,V)$ for the doped device as a function of voltage can be calculated by,

$$\psi(x,V) = -k_\text{B} T \ln\left(\frac{2\pi^2 \varepsilon_\text{r} \varepsilon_0 k_\text{B} T}{q^2 L^2 N_\text{C}}\left[\cos^2\left\{\frac{\pi x}{L} - \frac{\pi}{2}\right\}\right]^{-1} + \frac{3}{4}\frac{\varepsilon_\text{r}\varepsilon_0}{qN_\text{C}}\frac{V}{L^{3/2}}x^{-1/2} + \frac{n_\text{D}(x)}{N_\text{C}}\right). \tag{16}$$

With eq. 15 and 16, we now have excellent descriptions of the electron density and conduction-band edge of both undoped and doped electron-only devices.

### E. Full *J-V* descriptions of doped single-carrier device

Given the accurate description of the charge-carrier density, we are now capable of deriving a full description of the current density for a doped single-carrier device. Similar to how eq. 14 was calculated, we can now take the harmonic mean of eq. 15 to obtain an expression for the total drift current,

$$\langle n \rangle = \frac{4\pi^2 \varepsilon_\text{r} \varepsilon_0 k_\text{B} T}{q^2 L^2} + \frac{9 \varepsilon_\text{r} \varepsilon_0 V}{8 q L^2} + n_\text{D}. \tag{17}$$

A comparison between the numerically calculated $\langle n \rangle$ and eq. 17 for a 100 nm, 1 μm and 10 μm devices, evaluated at 0 V, is shown in **Fig. 5**. Excellent agreement is found for all three cases, namely i) when doping is not affecting the total electron density, ii) in the intermediate regime where doping mainly affects the middle of the device while the interfaces are affected by the background charge-carrier density, and iii) in the high doping limit.

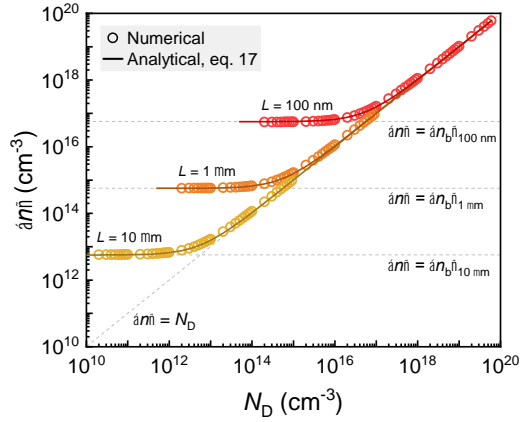

**Figure 5** – Numerically and analytically (eq. 17) calculated values for $\langle n \rangle$, for three electron-only devices with semiconductor thicknesses of 100 nm (red), 1 μm (orange) and 10 μm (yellow), as a function of $N_D$. The values for $\langle n \rangle = N_D$ and $\langle n \rangle = \langle n_b \rangle$ are shown as dashed lines.

Numerically calculated *J-V* curves of a 100 nm device with an increased density of doping, along with the corresponding slope-voltage (*m-V*) curves, $m = d \log J / d \log V$, are shown in **Fig. 6a**. When superimposing the numerical *J-V* curves with curves calculated by taking the sum of the Mott-Gurney law (eq. 2) and Ohm's law (eq. 3), a poor fit is obtained in the low voltage regime for low values of $N_D$. This poor fit is due to the omission of $n_b$, which is evident from the fact that when $n_D$ increases, the fit gradually improves since $n_b$ can now be ignored. While taking the sum of the Mott-Gurney law and Ohm's law gives a poor fit for low values of $n_D$, this sum is a good approximation in the high doping limit.

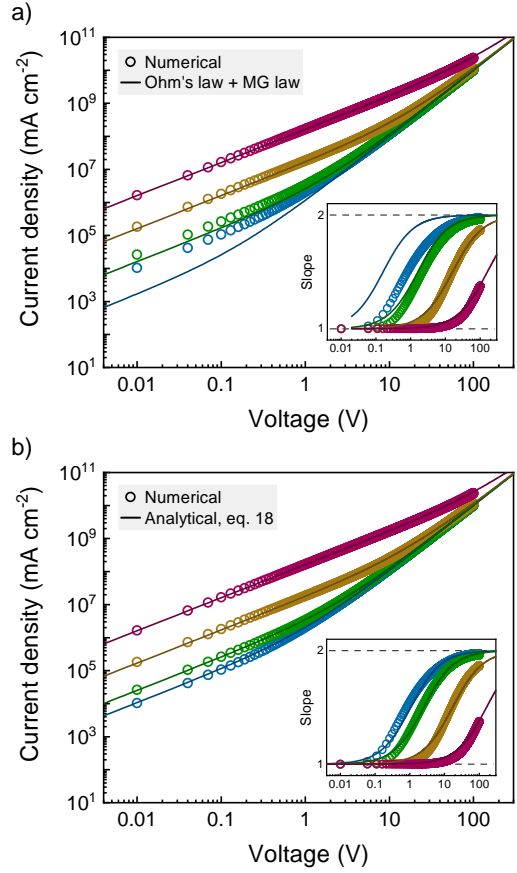

**Figure 6** – Numerically calculated *J-V* curves of 100 nm thick single-carrier devices where $N_D$ is varied from $10^{16}$ to $10^{19}$ cm$^{-3}$. a) The solid lines are calculated by taking the sum of the Mott-Gurney law (eq. 2) and Ohm's law (eq. 3). b) The solid lines are calculated with eq. 18. The corresponding *m-V* curves are shown as insets.

Similar to how the full *J-V* characteristics of an undoped single-carrier device can be modelled using a sum of eq. 1 and the Mott-Gurney law, a full description of the *J-V* relationship of a doped device can now be derived by insertion of eq. 17 into $J = q\mu_n \langle n \rangle V/L$,

$$J = q\mu_n \left(\frac{4\pi^2 \varepsilon_r \varepsilon_0 k_B T}{q^2 L^2} + n_D\right)\frac{V}{L} + \frac{9}{8}\mu_n \varepsilon_r \varepsilon_0 \frac{V^2}{L^3}. \qquad (18)$$

As shown in **Fig. 6b**, eq. 18 describes both the low voltage regime and the Mott-Gurney regime regardless of whether the semiconductor is undoped, lightly doped or doped close to degeneracy. In fact, an excellent agreement is found both for the overall magnitude and slope of the *J-V* curve, as seen from the inset in **Fig. 6b**. Equation 18 yields the sum of eq. 1 and the Mott-Gurney law when $n_D = 0$, and is therefore a more general description of the *J-V* characteristics of a single-carrier device. It should be noted that when the intrinsic charge-carrier density contributes to the current, $n_i$ can simply be added to eq. 16 ($n = n_b + n_D + n_i$). Since such an excellent agreement is found between eq. 18 and the numerically calculated *J-V* curves regardless of the value of $N_D$, eq. 18 can be used to fit SCLC data to obtain information about $\mu_n$ and the doping density simultaneously.

It is important to note that eq. 18 was derived assuming perfect ohmic contacts, $q\phi_{inj}(0) = q\phi_{inj}(L) = 0$ eV, and no influences from external resistances. Perfect ohmic contacts can be difficult to achieve due to mismatches between contact work functions and the transport levels of the probed semiconductor, and surface states on the semiconductor can also give rise to Fermi-level pinning, leading to injection issues. Additionally, if contacts are used that are much less conductive than the probed semiconductor, a series resistance can mask the device current. The effect of non-ohmic contacts and external resistances are discussed in the supplementary material, and we find that both of these effects diminish when thick devices are measured.

## F. Transition from undoped to doped conduction

Since the charge-carrier density at the conductor-semiconductor boundaries will always be dominated by $n_\text{b}$, and the electron density increases towards the middle of the device when you add donors, a requirement for the magnitude of the doping density, that must be added before it affects the device, can now be defined. From eq. 15 evaluated at $V = 0$ V, we define a condition for how large $N_\text{D}$ (and hence $n_\text{D}$) would have to be before affecting the overall electron density and hence the $J$-$V$ curves. This can be written as,

$$N_\text{D} > \langle n_\text{b} \rangle, \quad (19)$$

As the thickness of the semiconductor increases, a lower doping density can be detected from the $J$-$V$ curves, meaning that the thicker the single-carrier device is, the more sensitive to doping it will be. When measuring lightly doped semiconductors with SCLC, one should therefore always aim at measuring relatively thick devices following the condition described by eq. 19, simultaneously diminishing the effects from injection barriers and external resistances.

Finally, an additional tool can be derived by considering the cross-over voltage between the linear regime and the Mott-Gurney regime,

$$V_\text{X} = \frac{32\pi^2}{9}\frac{k_\text{B}T}{q} + \frac{8}{9}\frac{qn_\text{D}L^2}{\varepsilon_\text{r}\varepsilon_0}. \quad (20)$$

In the absence of doping, $V_\text{X}$ will take a value of ~0.9 V at 300 K; however, in the case where $N_\text{D} > \langle n_\text{b} \rangle$, a shift in $V_\text{X}$ will be observed according to eq. 20. Equations 18 and 20 can therefore be used in combination as reliable tools to characterize doping from SCLC data and to obtain meaningful values for the charge-transport characteristics.

## IV. CONCLUSIONS

We have here shown that while it is sufficient to take the sum of Ohm's law (eq. 3) and the Mott-Gurney law (eq. 2) when describing $J$-$V$ curves obtained from a single-carrier device containing a highly doped semiconductor, this is not sufficient when describing a device in which the semiconductor is lightly doped. To that end, we have derived a series of analytical expressions that can describe the charge-carrier density (eq. 15) and conduction-band edge (eq. 16), and hence the current density of a single-carrier device (eq. 18), regardless of whether the semiconductor is undoped, lightly doped or heavily doped. We have given a condition for how doped the semiconductor must be before the $J$-$V$ curves are significantly affected by doping (eq. 19), and we have shown that to model $J$-$V$ curves obtained from a lightly doped semiconductor with accuracy, both the background charge-carrier density and the doping density must both be taken into account. The analytical expressions presented herein can be fitted to SCLC data to yield information about charge-carrier mobility and the doping density simultaneously.

## SUPPLEMENTARY MATERIAL

See the supplementary material for additional discussions about the numerical simulations and the influence from non-ohmic contacts and external resistances.

## ACKNOWLEDGEMENTS

J.A.R. would like to thank Dr. André D. Taylor for allowing for the opportunity to publish this paper, Dr. Allison Kalpakci for reviewing the manuscript prior to submission, and Mr. Toke W. Fritzemeier and Dr. Morten Kjaergaard for valuable feedback.

## DATA AVAILABILITY STATEMENT

The data that support the findings of this study are available from the corresponding author upon reasonable request.